\documentclass{article}
\usepackage{spconf,amsmath,graphicx}
\usepackage{xcolor, bm, multirow}
\usepackage{amssymb}
\usepackage{subcaption}
\usepackage{xcolor}
\usepackage[colorlinks=true, allcolors=blue]{hyperref}

\title{Attention based Dual-Branch Complex Feature Fusion Network for Hyperspectral Image Classification}

 \name{Mohammed Q. Alkhatib $^{*}$, Mina Al-Saad, Nour Aburaed, M. Sami Zitouni, Hussain Al Ahmad}

\address{College of Engineering and IT, University of Dubai, Dubai, UAE\\
$^{*}$ mqalkhatib@ieee.org}

\begin{document}
%
\maketitle
\begin{abstract}
This research work presents a novel dual-branch model for hyperspectral image classification that combines two streams: one for processing standard hyperspectral patches using Real-Valued Neural Network (RVNN) and the other for processing their corresponding Fourier transforms using Complex-Valued Neural Network (CVNN). The proposed model is evaluated on the Pavia University and Salinas datasets. Results show that the proposed model outperforms state-of-the-art methods in terms of overall accuracy, average accuracy, and Kappa. Through the incorporation of Fourier transforms in the second stream, the model is able to extract frequency information, which complements the spatial information extracted by the first stream. The combination of these two streams improves the overall performance of the model. Furthermore, to enhance the model performance, the Squeeze and Excitation (SE) mechanism has been utilized. Experimental evidence show that SE block improves the models overall accuracy by almost 1\%. The project can be accessed at \url{ https://github.com/mqalkhatib/Real_Complex_Classification}
\end{abstract}
\begin{keywords}
HSI classification, Complex-Valued CNNs, Fourier Transform, Dual Branch.
\end{keywords}
\section{Introduction}
\vspace{-1em}
Hyperspectral image (HSI) data has been available since the 1980's \cite{qu2022review}. It can  provide massive amounts of spectral and spatial information in hundreds of narrow contiguous spectral bands ranging from visible to infrared wavelengths, which in turn makes challenging fine-grained remote sensing tasks possible. At present, HSI classification has become an interesting topic in the field of hyperspectral remote sensing, as HSI is used in a wide variety of earth observation applications, such as land cover mapping and environmental monitoring. HSI classification accuracy is an important criteria for such applications. In order to achieve precise and accurate classification results, it is essential to obtain effective spatial and spectral features. Traditional Machine Learning (ML) techniques were first introduced in the early research studies of HSI classification \cite {melgani2004classification,song2016hyperspectral}. These techniques use pixel-wise classification methods to classify each pixel in HSI data based on spectral information. These approaches mainly rely on the spectral characteristics of the features \cite{dasi2020hyperspectral}, which in turn makes it more challenging to obtain a good performance, especially with the presence of spectral variability and abundant mixed pixels \cite{ma2022hyperspectral}. Thus, spectral and spatial data could be merged to enhance HSI classification. 
\\
With the rapid development in Deep Learning (DL) technology, researchers in the field of computer vision began to use DL approaches, particularly Deep Convolutional Neural Networks (DCNNs) for classifying HSI features as they have shown superiority over other traditional classification methods \cite{alkhatib2023tri,ahmad2020fast}. The work of \cite{hu2015deep} presented 1D-CNN method to extract the spectral features. However, spectral information is not enough to get accurate classification results. To tackle this problem, researchers in \cite{makantasis2015deep} presented 2D-CNN approach to learn the spatial information for HSI classification. However, these methods did not exploit the full advantage of the 3D nature of HSI cube. Thus, researchers in \cite{hamida20183} proposed a 3D-CNN approach in which both spectral and spatial information are utilized to boost the classification performance. Roy et al. developed Hybrid Spectral CNN (HybridSN) \cite{roy2019hybridsn}, by fusing 2D-CNN with 3D-CNN, where 3D-CNN facilitates the joint spatial–spectral feature representation at the early stage, 2D-CNN is used for extracting more abstract-level spatial representation. The literature is rich with other similar examples \cite{alkhatib2023tri,yu2020simplified}. 
\begin{figure*} [t]
   \begin{center}
   \includegraphics[width=0.9\linewidth]{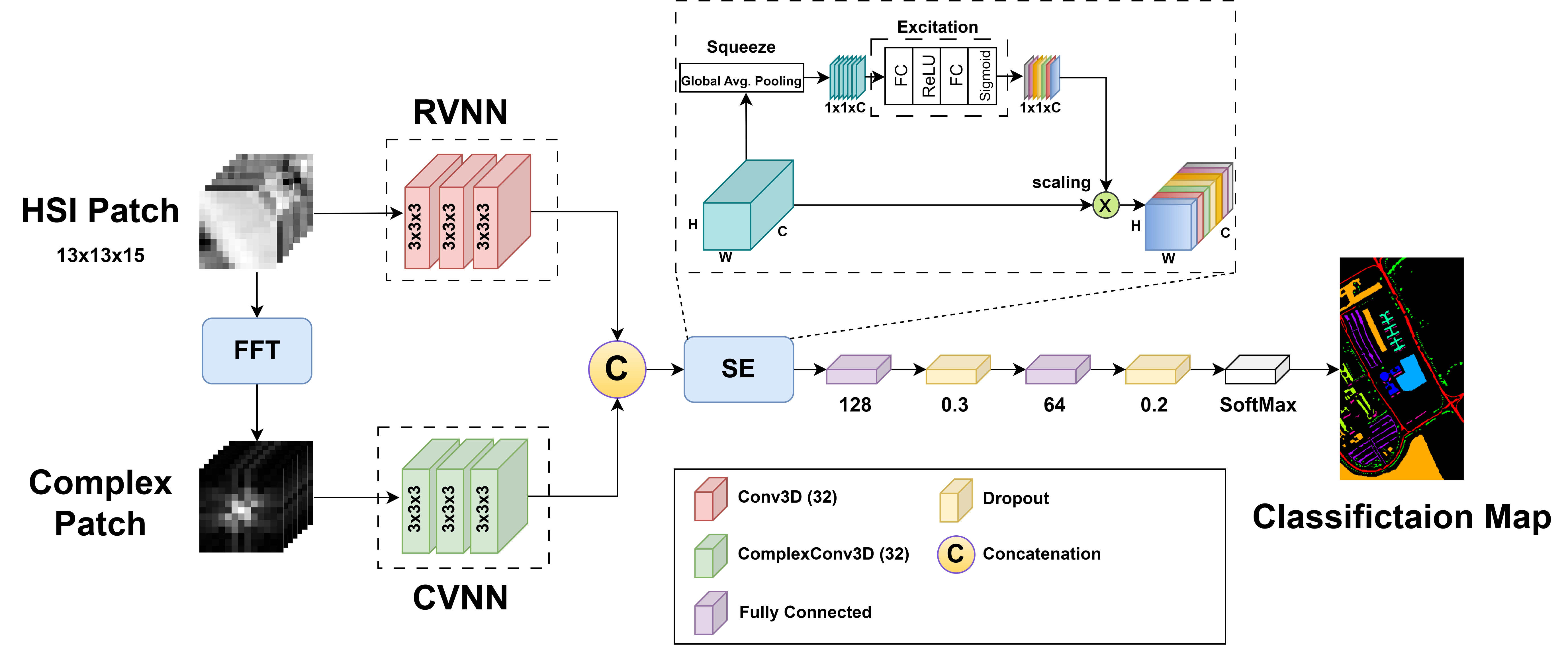}
   \end{center}
   \caption[example] 
   { \label{fig:flowchart}Overall Architecture of the proposed model for HSI classification.}
   \end{figure*} 
In the recent years, in order to further improve the classification performance, many research works widely investigated the channel attention mechanism and applied it to the field of HSI classification. One example is Squeeze and Excitation (SE) mechanism that models the interdependencies between the channels, which can increase the feature
representation capability of the model and improve the overall performance. For instance, Wang et al. \cite{wang2019spatial} introduced SE block into residual networks to improve the characterization of the spatial-spectral information contained in the HSI data.
Some research studies suggest that processing images in the frequency domain offers direct access to high and
low frequency components, which in turn can improve feature extraction process.
\\
\indent Nowadays, Complex-Valued Neural Networks (CVNNs) \cite{barrachina2021complex} are introduced as a variant of deep neural networks, where the properties of natural images can be captured easily using complex-valued features. Complex numbers have proven to be efficient for handling images. For example, the Fourier transform is complex valued, and have been considered in a neural network related context. A recent study combines Fourier transform and CVNN for hyperspectral single-image super resolution \cite{aburaed2023complex}. CVNNs exhibit great performance in processing complex-valued data such as radar and medical data \cite{lee2022complex}. This type of networks has not been yet investigated for HSI classification tasks.\\
\indent Based on the mentioned ideas above, this paper proposes a dual-branch network for HSI classification that combines two streams; one for processing standard hyperspectral patches and the other for processing their corresponding Fourier transforms. Attention mechanism is also introduced in our proposed methodology. Moreover, the proposed model is trained and tested using two benchmark HSI datasets, Pavia University (PU) and Salinas (SA). The rest of the paper is organized as follows: Section \ref{sec:model} explains the architecture and building blocks of the proposed model, experimental results and comparisons against state-of-the-art models are discussed in Section \ref{sec:results}, and finally, Section \ref{sec:con} summarizes the paper and states the future direction of this research.

\section{NETWORK ARCHITECTURE} \label{sec:model}
\vspace{-1em}
Due to the fact that HSIs are real-valued, there is no much research going on implementing CVNNs for HSI classification tasks. To the best of our knowledge, this is the first study that incorporates Complex-Valued Convolutional Neural Networks (CV-CNNs) for HSI classification tasks. First, the hyperspectral image cube will undergo dimensionality reduction using Principal Component Analysis (PCA), disjoint image patches are then extracted. Fast Fourier Transform (FFT) is applied to the extracted patches in a Band-wise manner to obtain the Complex Patch. The original HSI patch will go through the RVNN stream while the complex patch will go through the CVNN stream. The extracted features from both sides are then fused using concatenation. The fused features will then pass through the SE block to enhance channel interdependencies and improve the model performance. Classification will be performed by a series of fully-connected and dropout layers. Fig. \ref{fig:flowchart} shows an overview of the proposed model.
\subsection{RVNN stream}

The RVNN stream consists of 3 layers of 3D-CNN to extract spatial-spectral features. 3D convolution performs computations over the height, width, and depth (bands) of each HSI cube instead of processing each channel individually \cite{aburaed20213d} and has been proven to be more suitable for HSI processing \cite{hamida20183}. 

\subsection{CVNN stream}
 CVNNs are a type of CNN which has all parameters (input, weight, biases and output values) are of complex type numbers. Compared to real-value ones, CVNNs have many advantages such as improved performance and they can learn faster and achieve better generalizations. CV-CNNs widely employed for Synthetic Aperture Radar (SAR) image interpretation \cite{zhang2017complex}. As mentioned earlier, 3D convolution operations are more suited for processing HSI than 2D. The complex value counterpart of a conventional real-valued 3D convolution, where each band of the HSI cube must be converted to the complex domain and convolved with a complex filter using Band-wise Fast Fourier Transform (FFT). 3D Complex Convolutional layer is defined as eq. \ref{3D-CVNNeq}. The CVNN stream uses 3 layers of CV-CNN to process the complex patches.

\footnotesize
\begin{equation}\label{3D-CVNNeq}
\ F_{(x,y,z)} = \!\mathbb{C} \textrm{ReLU}\!\left (\!\sum_{i=1}^{M}\!\sum_{j=1}^{M}\!\sum_{k=1}^{C} K_{(i,j,k)} X_{(x+i,y+j,z+k)} + \!b \right)\
\end{equation}
\normalsize

where $X$ is the complex HSI patch $[X=X_{r}+iX_{im}]$ , $K$ is complex filter $[K=K_{r}+iK_{im}]$ and $b$ is the complex bias. The complex-valued activation function $\mathbb{C} \textrm{ReLU(.)}$, simply consists in applying the well known real-valued function ReLU(.) to both the real and imaginary parts separately so that $\mathbb{C} \textrm{ReLU(z) = ReLU(Re(z)) + }i\textrm{ReLU(Im(z))} $
.

\subsection{Attention Block}
In the SE process \cite{zhang2022sem}, For
any given transformation (e.g. convolution) $F_{tr}$ for mapping the input $X$ to the feature maps $u_{c}$ where $u_{c}\in \mathbb{R}^{H\times W\times C}$ , $u_{c}$ represents the c-th
$H \times W $matrix in $u$, and the subscript $c$ represents the number of channels. In Squeeze procedure ${F_{sq}(.)}$, it uses global average
pooling to convert the input of $H \times  W \times C$ into the output of $1 \times 1 \times C$.The squeeze function is defined as 
\begin{equation}
\label{Squeeze}
\ z_{c}=F_{sq(u_{c})} = \frac{1}{H\times W}\sum_{i=1}^{H}\!\sum_{j=1}^{W} u_{c}(i,j)\
\end{equation}
It can be seen from the equation that input data eventually become a column vector in the squeeze process. The length of this vector is the same as the number of channels. After that, the excitation operation is performed to learn the importance of each feature automatically which in turn amplify the features that have a greater impact on the classification results while suppressing useless features.The excitation function can be expressed as 
\begin{equation}
\label{excitation}
\ s= F_{ex(z,W)} = \sigma(g(z,W))=\sigma(W_{2} 
 ReLU(W_{1}z))\
\end{equation}
where $\sigma$ is the Sigmoid activation function, $W_{1}\in \mathbb {R}^{\frac{C}{r}\times C}$ and $W_{2}\in \mathbb {R}^{ C\times \frac{C}{r}}$are the two fully connected layers, $W_{1}$ is the dimensionality reduction layer with a dimensionality reduction ratio of $r$. $W_{2}$ is the proportionally identical data-dimensionality increase layer. $z$ is the output from previous squeeze layer, $W_{1} z$ is a fully connected layer operation. The Sigmoid function processes the final output of the excitation process to a value between zero and one. After the squeeze and excitation steps, we get a attention
vector $s$, which we can use to adjust the channels of $u$.

\section{Experiments and Analysis} 
\begin{figure} [t]
   \begin{center}
   \includegraphics[width= 1\linewidth]{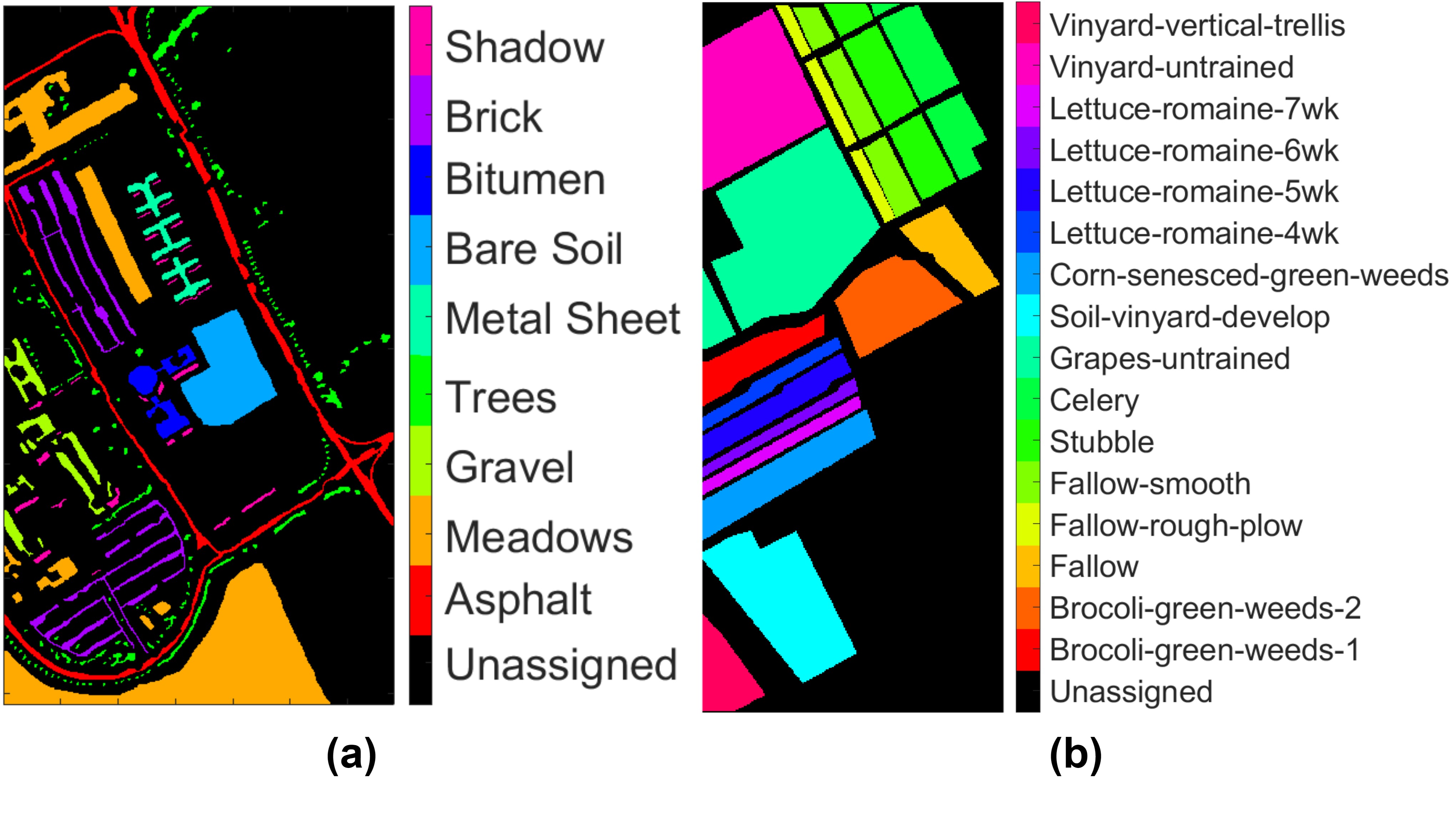}
   \end{center}
   \caption{ \label{fig:datasets}Reference Data: (a) PU; (b) SA}
   \end{figure}

\vspace{-1em}
\label{sec:results}
To demonstrate the performance of the proposed approach shown in Figure \ref{fig:flowchart}, the proposed methodology is compared with 2D-CNN \cite{makantasis2015deep}, 3D-CNN \cite{hamida20183}, PMI-CNN \cite{zhong2022hyperspectral} and HybridSN \cite{roy2019hybridsn}. Also, to assess the impact of the Attention SE block on model performance, experiments are conducted by removing the block and evaluating the model's performance without it. The Overall Accuracy (OA), the Average Accuracy (AA), and the Kappa statistic (Kappa) are reported to evaluate the performance of the proposed model. The classification accuracy of each class is also provided. The experiments were conducted and repeated 10 times. For all algorithms, only the classification results with the highest accuracy in 10 trials are recorded. For OA, AA, and Kappa the average and standard deviation of all 10 trials are recorded. The proposed model is evaluated using two widely used hyperspectral datasets: PU and SA. Figure \ref{fig:datasets} shows the reference data for both datasets. Full description is available in \cite{dataURL}

\begin{figure} [t]
   \centering
   \includegraphics[width= 1\linewidth]{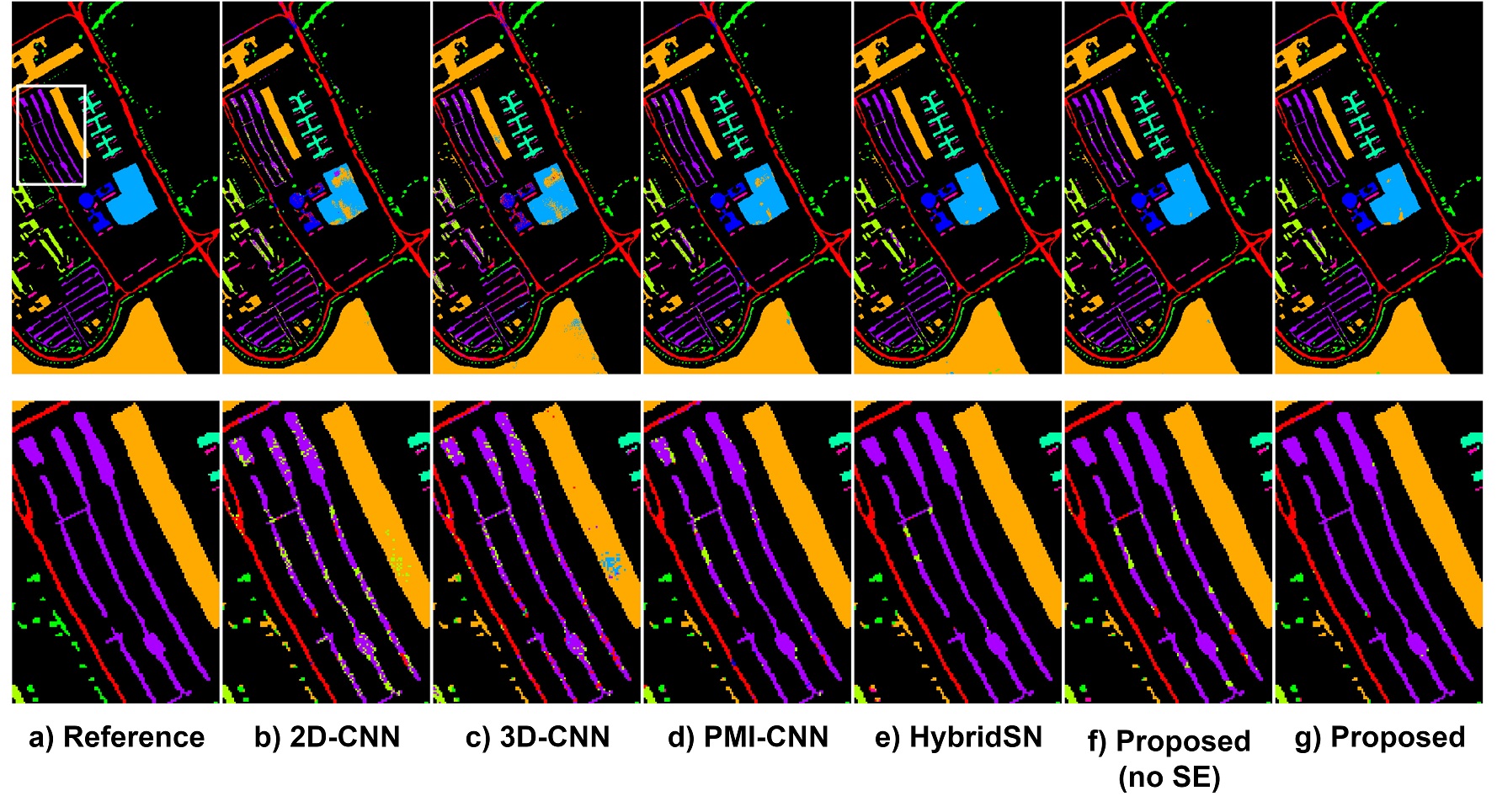}   
   \caption{ \label{fig:classification_PU} Classification maps of PU Dataset. (a) Reference Data; (b) 2D-CNN; (c) 3D-CNN; (d) PMI-CNN (e) HybridSN; (f) Proposed (no SE); (g) Proposed}
   \end{figure}

\begin{figure} [t]
   \centering
   \includegraphics[width= 1\linewidth]{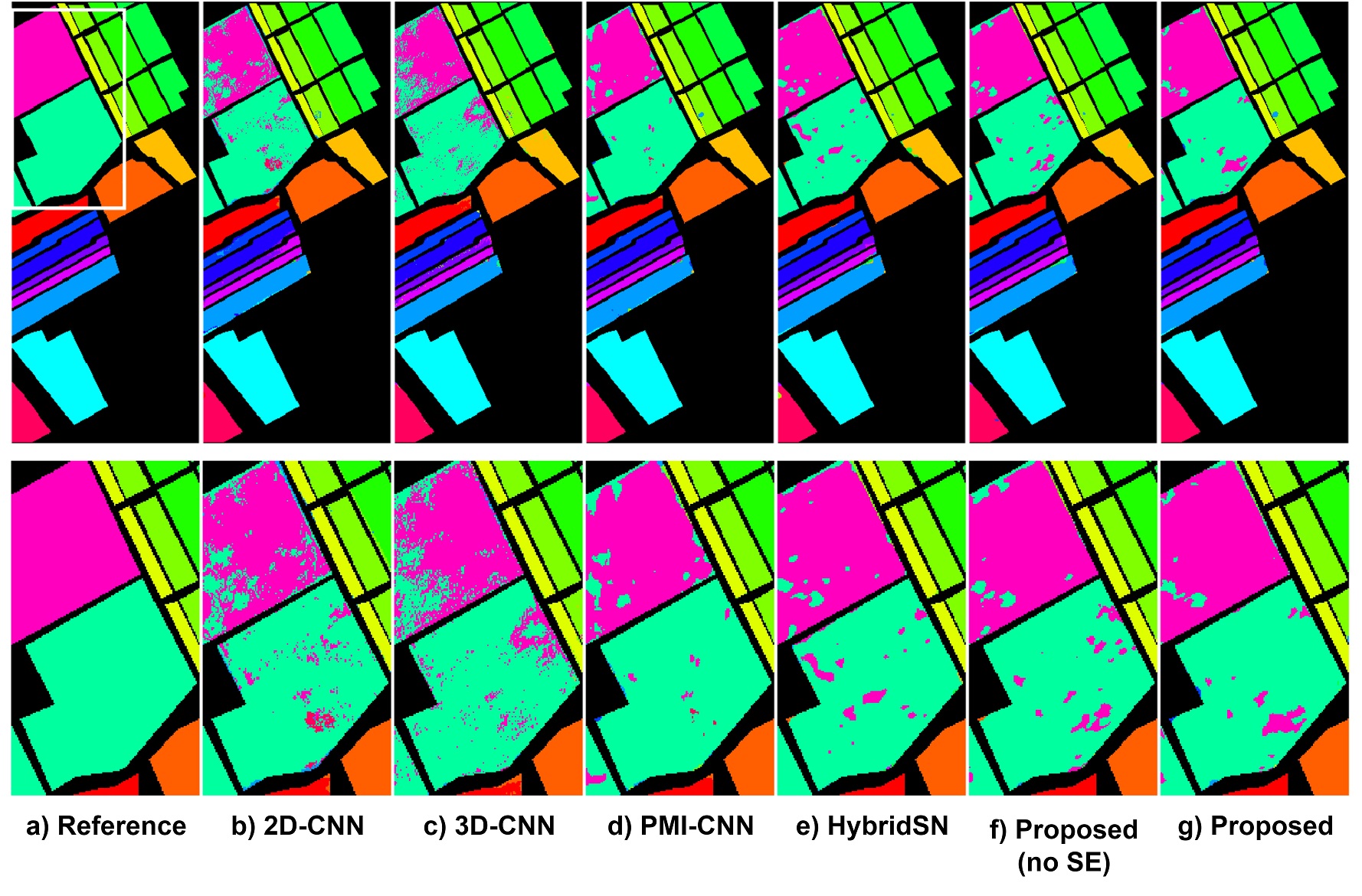}
   \caption{  \label{fig:classification_SA} Classification maps of SA Dataset. (a) Reference Data; (b) 2D-CNN; (c) 3D-CNN; (d) PMI-CNN (e) HybridSN; (f) Proposed (no SE); (g) Proposed}
   \end{figure}

\indent For both datasets, image patches were randomly divided into 1\% for training, and the remaining 99\% are used for testing and evaluation, we used  a patch size of $13\times13$ and the number of principal components was set to 15, multiple attempts were made through trials to discover the most suitable values, leading to the determination of these optimal choices. The model is trained for 100 epochs with batch size of 16. During the model training, early stopping strategy is adopted. Specifically, if the model's performance did not improve over 10 consecutive epochs, the training process was terminated and the model was restored to its best weights. The optimization algorithm is Adam. The learning rate is set to $10^{-3}$. The loss function is categorical cross-entropy. All models are implemented using Python, Keras framework that runs with TensorFlow back-end. To ensure a fair comparison, all models were trained under the same conditions.\\ 
\indent Tables \ref{PU_Qualitativeres} and \ref{SA_Qualitativeres} list the classification accuracies (including classification accuracy of each class, OA, AA, and Kappa) for both datasets. It can be seen that the proposed model can achieve superior results than other classification models. In comparison with HybridSN, the proposed model accuracy increases by almost 1.3\% on the PU dataset. Also, by utilizing the attention mechanism, the model performance increases by almost 1\% when compared to its no-SE version of the model. Besides, the results show less variability or fluctuation compared to other models. This indicates that the proposed model is performing well and producing reliable and accurate predictions.\\
\indent To conduct the ablation study, we evaluated the data separately in each stream. Specifically, we used the standard real-type data in the RVNN stream and the FFT data in the CVNN stream. The outcomes revealed that the RVNN stream achieved an accuracy of 95.1\% for Pavia and 95.53\% for Salinas. On the other hand, the CVNN stream achieved scores of 94.74\% for Pavia and 95.02\% for Salinas. However, when the features were fused together, the model's performance improved, resulting in an increase in accuracies of more than 1\% for both datasets, as reported in Tables \ref{PU_Qualitativeres} and \ref{SA_Qualitativeres}.\\
\indent Figures \ref{fig:classification_PU} and \ref{fig:classification_SA} shows the classification results for both datasets. The upper row shows the full class map while the bottom row shows an enlarged area (Marked by the white box on the full reference map). The proposed model with attention shows the closest visual quality to the ground truth when compared to other methods.

\begin{table}[t]
\centering
\caption{Classification performance of different methods for the PU dataset. Bold indicates the best result}
\label{PU_Qualitativeres}
\resizebox{3.4in}{!}{
\begin{tabular}{ccccccc}
\hline
Class       & 2D-CNN          & 3D-CNN          & PMI-CNN         & HybridSN        & \begin{tabular}[c]{@{}c@{}}Proposed\\ (no SE)\end{tabular} & Proposed            \\ \hline
Asphalt          & 93.37           & 89.51           & 92.98           & 99.54           & \textbf{99.74}                                             & 98.51               \\
Meadows          & 99.71           & 98.53           & 99.63           & 99.65           & 99.91                                                      & \textbf{99.92}      \\
Gravel          & 80.70           & 53.32           & 82.05           & 89.46           & 87.44                                                      & \textbf{89.70}      \\
Trees          & 91.26           & 79.62           & 92.71           & 89.12           & 90.01                                                      & \textbf{92.82}      \\
Metal Sheet          & \textbf{100.00} & \textbf{100.00} & \textbf{100.00} & \textbf{100.00} & \textbf{100.00}                                            & \textbf{100.00}     \\
Bare Soil          & 77.06           & 83.31           & 94.01           & 98.21           & \textbf{99.48}                                             & 96.45               \\
Bitumen          & 83.30           & 67.50           & 98.18           & 98.63           & 98.25                                                      & \textbf{99.32}      \\
Brick          & 80.71           & 79.09           & 91.55           & 95.42           & 91.14                                                      & \textbf{96.63}      \\
Shadow          & 76.20           & 72.36           & \textbf{99.89}  & 94.02           & 97.87                                                      & 95.62               \\ \hline
OA (\%)     & 88.27$\pm$1.93      & 86.16$\pm$1.5       & 94.00$\pm$1.54      & 95.62$\pm$1.19      & 95.96$\pm$0.91                                                 & \textbf{96.99$\pm$0.4}  \\
AA (\%)     & 81.94$\pm$4.29      & 78.32$\pm$2.51      & 91.51$\pm$2.25      & 93.27$\pm$2.07      & 93.40$\pm$1.43                                                 & \textbf{95.92$\pm$0.42} \\
Kappa x 100 & 84.25$\pm$2.6       & 81.36$\pm$2.05      & 91.97$\pm$2.11      & 94.17$\pm$1.58      & 94.62$\pm$1.23                                                 & \textbf{96.00$\pm$0.54} \\ \hline
\end{tabular}}
\end{table}
\begin{table}[t]
\centering
\caption{Classification performance of different methods for the SA dataset. Bold indicates the best result}
\label{SA_Qualitativeres}
\resizebox{3.4in}{!}{
\begin{tabular}{ccccccc}
\hline
Class       & 2D-CNN          & 3D-CNN          & PMI-CNN         & HybridSN        & \begin{tabular}[c]{@{}c@{}}Proposed\\ (no SE)\end{tabular} & Proposed            \\ \hline
Brocoli-green-weeds-1          & 98.74           & 96.63           & 99.40           & \textbf{100.00} & \textbf{100.00}                                            & \textbf{100.00}     \\
Brocoli-green-weeds-2          & \textbf{100.00} & \textbf{100.00} & \textbf{100.00} & \textbf{100.00} & \textbf{100.00}                                            & \textbf{100.00}     \\
Fallow           & 99.90           & 97.60           & 99.90           & 97.29           & 98.47                                                      & \textbf{100.00}     \\
Fallow-rough-plow          & 96.59           & 93.62           & 98.99           & 98.62           & 99.86                                                      & \textbf{99.93}      \\
Fallow-smooth          & 98.83           & 97.74           & 98.83           & 99.02           & \textbf{99.43}                                             & 98.00               \\
Stubble          & 95.28           & \textbf{100.00} & \textbf{100.00} & \textbf{100.00} & 99.97                                                      & \textbf{100.00}     \\
Celery          & 99.87           & 98.25           & 99.60           & 99.92           & 99.89                                                      & \textbf{99.94}      \\
Grapes-untrained           & 93.39           & 91.32           & \textbf{97.48}  & 95.04           & 94.60                                                      & 94.67               \\
Soil-vinyard-develop          & \textbf{100.00} & \textbf{100.00} & \textbf{100.00} & \textbf{100.00} & \textbf{100.00}                                            & \textbf{100.00}     \\
Corn-senesced-green-weeds         & 94.61           & 98.21           & 95.59           & 97.35           & 95.96                                                      & \textbf{98.24}      \\
Lettuce-romaine-4wk         & 88.65           & 95.08           & 98.39           & 96.50           & 97.82                                                      & \textbf{98.96}      \\
Lettuce-romaine-5wk         & 96.80           & 98.32           & 98.48           & \textbf{100.00} & 99.79                                                      & 99.95               \\
Lettuce-romaine-6wk         & 74.42           & 87.21           & 98.57           & 95.81           & 95.70                                                      & \textbf{99.23}      \\
Lettuce-romaine-7wk         & 98.49           & 93.58           & 99.34           & 98.02           & \textbf{100.00}                                            & 95.00               \\
Vinyard-untrained           & 78.65           & 75.14           & 88.85           & 92.83           & 92.58                                                      & \textbf{94.39}      \\
Vinyard-vertical-trellis          & 99.39           & 99.27           & 99.22           & 98.10           & 99.16                                                      & \textbf{99.66}      \\ \hline 
OA (\%)     & 90.98$\pm$1.26      & 92.69$\pm$0.68      & 94.83$\pm$1.2       & 96.05$\pm$1.10      & 96.10$\pm$1.07                                                 & \textbf{97.15$\pm$0.48} \\
AA (\%)     & 91.36$\pm$1.76      & 93.83$\pm$1.13      & 96.80$\pm$0.72      & 96.38$\pm$1.77      & 97.07$\pm$0.76                                                 & \textbf{97.8$\pm$0.97}  \\
Kappa x 100 & 90.95$\pm$1.41      & 91.86$\pm$0.75      & 94.24$\pm$1.33      & 95.60$\pm$1.22      & 95.66$\pm$1.18                                                 & \textbf{96.83$\pm$0.54} \\ \hline
\end{tabular}}
\end{table}

\vspace{-1em}

\vspace{-0.5em} 
\section{Conclusions}\label{sec:con}
\vspace{-1em}
In this paper, an end-to-end framework dual-branch attention-based feature fusion network for HSI classification is proposed. Our network consists of two streams: RVNN for processing standard hyperspectral patches and CVNN for processing their corresponding Fourier transforms. The SE block is also utilized as an attention mechanism within our architecture. FFT is used to transfer the real HSI to its complex counterpart. The network is trained, tested, and evaluated using PU and SA datasets. Extensive experimental results demonstrate that our proposed framework can achieve a decent performance with only 1\% of the data is used for training. Our approach with attention outperforms other models quantitatively and qualitatively. For future work, 
more experiments will be conducted to explore the use of other types of transform-based features, such as wavelet transforms, to further enhance the feature extraction capabilities of the model.

\bibliographystyle{IEEEbib}
\bibliography{strings,refs}

\begin{thebibliography}{10}

\bibitem{qu2022review}
Shenming Qu, Xiang Li, and Zhihua Gan,
\newblock ``A review of hyperspectral image classification based on joint
  spatial-spectral features,''
\newblock in {\em Journal of Physics: Conference Series}. IOP Publishing, 2022,
  vol. 2203, p. 012040.

\bibitem{melgani2004classification}
Farid Melgani and Lorenzo Bruzzone,
\newblock ``Classification of hyperspectral remote sensing images with support
  vector machines,''
\newblock {\em IEEE Transactions on geoscience and remote sensing}, vol. 42,
  no. 8, pp. 1778--1790, 2004.

\bibitem{song2016hyperspectral}
Weiwei Song, Shutao Li, Xudong Kang, and Kunshan Huang,
\newblock ``Hyperspectral image classification based on knn sparse
  representation,''
\newblock in {\em 2016 IEEE international geoscience and remote sensing
  symposium (IGARSS)}. IEEE, 2016, pp. 2411--2414.

\bibitem{dasi2020hyperspectral}
Syamala Dasi, Deekshitha Peeka, Reshma~Begum Mohammed, and BLN~Phaneendra
  Kumar,
\newblock ``Hyperspectral image classification using machine learning
  approaches,''
\newblock in {\em 2020 4th International Conference on Intelligent Computing
  and Control Systems (ICICCS)}. IEEE, 2020, pp. 444--448.

\bibitem{ma2022hyperspectral}
Chen Ma, Junjun Jiang, Huayi Li, Xiaoguang Mei, and Chengchao Bai,
\newblock ``Hyperspectral image classification via spectral pooling and hybrid
  transformer,''
\newblock {\em Remote Sensing}, vol. 14, no. 19, pp. 4732, 2022.

\bibitem{alkhatib2023tri}
Mohammed~Q Alkhatib, Mina Al-Saad, Nour Aburaed, Saeed Almansoori, Jaime
  Zabalza, Stephen Marshall, and Hussain Al-Ahmad,
\newblock ``Tri-cnn: a three branch model for hyperspectral image
  classification,''
\newblock {\em Remote Sensing}, vol. 15, no. 2, pp. 316, 2023.

\bibitem{ahmad2020fast}
Muhammad Ahmad, Adil~Mehmood Khan, Manuel Mazzara, Salvatore Distefano, Mohsin
  Ali, and Muhammad~Shahzad Sarfraz,
\newblock ``A fast and compact 3-d cnn for hyperspectral image
  classification,''
\newblock {\em IEEE Geoscience and Remote Sensing Letters}, vol. 19, pp. 1--5,
  2020.

\bibitem{hu2015deep}
Wei Hu, Yangyu Huang, Li~Wei, Fan Zhang, and Hengchao Li,
\newblock ``Deep convolutional neural networks for hyperspectral image
  classification,''
\newblock {\em Journal of Sensors}, vol. 2015, pp. 1--12, 2015.

\bibitem{makantasis2015deep}
Konstantinos Makantasis, Konstantinos Karantzalos, Anastasios Doulamis, and
  Nikolaos Doulamis,
\newblock ``Deep supervised learning for hyperspectral data classification
  through convolutional neural networks,''
\newblock in {\em 2015 IEEE international geoscience and remote sensing
  symposium (IGARSS)}. IEEE, 2015, pp. 4959--4962.

\bibitem{hamida20183}
Amina~Ben Hamida, Alexandre Benoit, Patrick Lambert, and Chokri~Ben Amar,
\newblock ``3-d deep learning approach for remote sensing image
  classification,''
\newblock {\em IEEE Transactions on geoscience and remote sensing}, vol. 56,
  no. 8, pp. 4420--4434, 2018.

\bibitem{roy2019hybridsn}
Swalpa~Kumar Roy, Gopal Krishna, Shiv~Ram Dubey, and Bidyut~B Chaudhuri,
\newblock ``Hybridsn: Exploring 3-d--2-d cnn feature hierarchy for
  hyperspectral image classification,''
\newblock {\em IEEE Geoscience and Remote Sensing Letters}, vol. 17, no. 2, pp.
  277--281, 2019.

\bibitem{yu2020simplified}
Chunyan Yu, Rui Han, Meiping Song, Caiyu Liu, and Chein-I Chang,
\newblock ``A simplified 2d-3d cnn architecture for hyperspectral image
  classification based on spatial--spectral fusion,''
\newblock {\em IEEE Journal of Selected Topics in Applied Earth Observations
  and Remote Sensing}, vol. 13, pp. 2485--2501, 2020.

\bibitem{wang2019spatial}
Li~Wang, Jiangtao Peng, and Weiwei Sun,
\newblock ``Spatial--spectral squeeze-and-excitation residual network for
  hyperspectral image classification,''
\newblock {\em Remote Sensing}, vol. 11, no. 7, pp. 884, 2019.

\bibitem{barrachina2021complex}
J.~A. Barrachina, C.~Ren, C.~Morisseau, G.~Vieillard, and J.-P. Ovarlez,
\newblock ``Complex-valued vs. real-valued neural networks for classification
  perspectives: An example on non-circular data,''
\newblock in {\em ICASSP 2021 - 2021 IEEE International Conference on
  Acoustics, Speech and Signal Processing (ICASSP)}, 2021, pp. 2990--2994.

\bibitem{aburaed2023complex}
Nour Aburaed, Mohammed~Q Alkhatib, Stephen Marshall, Jaime Zabalza, and Hussain
  Al~Ahmad,
\newblock ``Complex-valued neural network for hyperspectral single image super
  resolution,''
\newblock in {\em Hyperspectral Imaging and Applications II}. SPIE, 2023, vol.
  12338, pp. 102--109.

\bibitem{lee2022complex}
ChiYan Lee, Hideyuki Hasegawa, and Shangce Gao,
\newblock ``Complex-valued neural networks: A comprehensive survey,''
\newblock {\em IEEE/CAA Journal of Automatica Sinica}, vol. 9, no. 8, pp.
  1406--1426, 2022.

\bibitem{aburaed20213d}
Nour Aburaed, Mohammed~Q Alkhatib, Stephen Marshall, Jaime Zabalza, and Hussain
  Al~Ahmad,
\newblock ``3d expansion of srcnn for spatial enhancement of hyperspectral
  remote sensing images,''
\newblock in {\em 2021 4th International Conference on Signal Processing and
  Information Security (ICSPIS)}. IEEE, 2021, pp. 9--12.

\bibitem{zhang2017complex}
Zhimian Zhang, Haipeng Wang, Feng Xu, and Ya-Qiu Jin,
\newblock ``Complex-valued convolutional neural network and its application in
  polarimetric sar image classification,''
\newblock {\em IEEE Transactions on Geoscience and Remote Sensing}, vol. 55,
  no. 12, pp. 7177--7188, 2017.

\bibitem{zhang2022sem}
Jiawei Zhang, Pingli Ma, Tao Jiang, Xin Zhao, Wenjun Tan, Jinghua Zhang,
  Shuojia Zou, Xinyu Huang, Marcin Grzegorzek, and Chen Li,
\newblock ``Sem-rcnn: a squeeze-and-excitation-based mask region convolutional
  neural network for multi-class environmental microorganism detection,''
\newblock {\em Applied Sciences}, vol. 12, no. 19, pp. 9902, 2022.

\bibitem{zhong2022hyperspectral}
Huan Zhong, Li~Li, Jiansi Ren, Wei Wu, and Ruoxiang Wang,
\newblock ``Hyperspectral image classification via parallel multi-input
  mechanism-based convolutional neural network,''
\newblock {\em Multimedia Tools and Applications}, vol. 81, no. 17, pp.
  24601--24626, 2022.

\bibitem{dataURL}
``Hyperspectral remote sensing scenes,''
  \url{https://www.ehu.eus/ccwintco/index.php/Hyperspectral_Remote_Sensing_Scenes},
\newblock Accessed: 22-Feb-2023.

\end{thebibliography}

\end{document}